\documentclass[12pt]{article}

\usepackage[pdftex]{graphicx}
\usepackage{bm}
\usepackage{xcolor}
\usepackage[colorlinks=true]{hyperref}

\usepackage{scicite}
\usepackage{times}

\newenvironment{sciabstract}{%
\begin{quote} \bf}
{\end{quote}}

\newcommand{\etal}{ {\it et al.}}

\renewcommand\thefigure{\arabic{figure}}  

\renewcommand\thetable{\arabic{table}}

\newcommand{\newc}{\newcommand}
 
\newc{\be}{\begin{equation}}
\newc{\ee}{\end{equation}}
\newc{\bfe}{\begin{floatequation}}
\newc{\efe}{\end{floatequation}}
\newc{\bea}{\begin{eqnarray}}
\newc{\eea}{\end{eqnarray}}
\newc{\ie}{{\it i.e.} }
\newc{\eg}{{\it e.g.} }
\newc{\etc}{{\it etc.} }
\newc{\ra}{\rightarrow}
\newc{\lra}{\leftrightarrow}
\newc{\lsim}{\buildrel{<}\over{\sim}}
\newc{\gsim}{\buildrel{>}\over{\sim}}

\newcommand{\orcid}[1]{\href{https://orcid.org/#1}{\includegraphics[width=8pt]{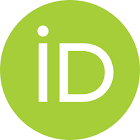}}}

\title{Information theory and Electron Spin Resonance dating}

\author{C. Tannous$^{1\ast}$ \orcid{0000-0002-9293-1763} and J. Gieraltowski$^{2}$ \\
\normalsize{$^{1}$ Universit\'e de  Brest, Lab-STICC, CNRS-UMR 6285, F-29200 Brest, FRANCE} \\
\normalsize{$^{2}$Laboratoire G\'eosciences-Oc\'ean (LGO), CNRS-UMR 6538, 29280 Plouzan\'e, FRANCE} \\
\\
\normalsize{$^\ast$To whom correspondence should be addressed; E-mail: tannous@univ-brest.fr}
}

\date{}

\begin{document} 


\baselineskip24pt


\maketitle

\begin{sciabstract}
Chronometric dating is becoming increasingly important in areas such as the Origin and
evolution of Life on Earth and other planets, Origin and evolution of the Earth and 
the Solar System...
Electron Spin Resonance (ESR) dating is based on exploiting effects of 
contamination by chemicals or ionizing radiation, on ancient matter through 
its absorption spectrum and lineshape.
Interpreting absorption spectra as probability density functions (pdf), 
we use the notion of Information Theory (IT) distance 
allowing us to position the measured lineshape with respect 
to standard limiting pdf's (Lorentzian and Gaussian). 
This paves the way to perform dating when several interaction patterns
between unpaired spins are present in geologic, planetary, meteorite or asteroid matter
namely classical-dipolar (for ancient times) and quantum-exchange-coupled (for recent times). 
In addition, accurate bounds to age are provided by IT
from the evaluation of distances with respect to the Lorentz and Gauss distributions.
Dating arbitrary periods of times~\cite{Anderson} and exploiting IT to introduce 
rigorous and accurate date values might have interesting far reaching implications not only
in Geophysics, Geochronology~\cite{Bahain}, Planetary Science but also in Mineralogy, Archaeology, 
Biology, Anthropology~\cite{Aitken}, Paleoanthropology~\cite{Taylor,Richter}...
\end{sciabstract}



\section*{Introduction}
ESR (or EPR for Electron Paramagnetic Resonance) absorption spectroscopy is a non-interfering 
versatile technique that allows to explore interaction between
unpaired spins and an applied magnetic field in condensed matter~\cite{Aitken2}. 
These unpaired spins pertain to electrons in general or electrons as well as 
holes in semiconducting materials.
CW-ESR (Continuous wave) method measures concentrations of paramagnetic centers  
and free radicals by shining
a sample with microwaves at a fixed frequency while simultaneously sweeping the magnetic field.
Pulsed-ESR, CW-ESR successor allowed to reduce
measurement time, increase sensitivity and resolution, better separate different interactions and
detect different types of spin relaxation mechanisms~\cite{Schweiger}.

ESR provides precious information about local structures  and dynamic 
processes of the paramagnetic centers within the sample under study. 

Different frequencies are used such as S band (3.5 GHz), X band (9.25 GHz), 
K band (20 GHz), Q band (35 GHz) and W band (95 GHz). Each frequency has its own advantages 
and drawbacks and increasing it might increase its sensitivity. The latter is the minimal
concentration of unpaired spins ESR can detect. For instance, going from the X to W band
may result into a 30,000 enhancement in sensitivity. Since X band static sensitivity
is around 10$^{12}$ spins/cm$^3$ this means the detection limit is decreased to 
3.33 $\times$ 10$^{7}$ spins/cm$^3$~\cite{Pan2002}. Other means for improving static and 
dynamic sensitivity (in spins/Gauss/$\sqrt{\rm Hz}$) involve reducing temperature or miniaturizing 
experimental parts such as the resonant cavity or
even employing cQED (circuit Quantum Electrodynamics) devices such as Josephson junctions,
and SQUIDS, potentially reaching single-spin sensitivity by using high
quality factor superconducting micro-resonators along with Josephson Parametric 
Amplifiers~\cite{Probst}.

Importance of interest in ESR absolute dating capabilities stemmed from the fact ionizing radiation 
($\alpha, \beta$ and $\gamma$) creates unpaired spins that might have extremely long lifetimes in 
certain materials~\cite{Skinner}. Ionizing radiation occurs since rocks, 
sediments~\footnote{Earth surface processes are best studied with optically stimulated 
luminescence (OSL) of sediments~\cite{Rhodes}.   
With OSL one can date deposits aged from one year to several hundred thousand years.},
minerals and deposits contain radioactive elements, such as potassium and isotopes produced
by the $^{238}$U (Uranium), $^{235}$U (Actinium) and $^{232}$Th (Thorium) decay series.

Accurate dosimetry estimation is needed for radiometric dating in order to estimate
how much any sample was exposed to ionizing radiation (or by extension to other processes) tying 
the dose absorbed by the sample to its age.  More precisely, the ESR signal intensity is proportional to 
the paleodose $D$ (total radiation dose) given by $D=\int_0^T D_R(t) \, dt$ where $D_R(t)$ is the
natural dose rate (in Grays/year) and $T$ the exposure time or estimated age~\cite{Ikeya,Jonas}.

Dating is important in geochronology and in Planetary Science since asteroids and meteorites contain
organic materials that might provide important clues for the sources of Life on Earth~\cite{Wagner}
and the formation of the Solar System.

It is to important to realize that ESR spectroscopy can handle radiometric dating as well 
as chemical dating that delivers information about various chemical processes, a mineral has been 
subjected to as in reference~\cite{Pan2002}.

While several other dating methodologies exist (see for instance Geyh and Schleicher~\cite{Geyh}
or Ikeya's book~\cite{Ikeya}),
the interest in ESR dating grew considerably after realizing its wide time range
since it spans from a few thousand years to several million  and even billions years
which is far beyond capabilities of $^{14}$C radio-carbon 
dating~\cite{Galli} (limited to about 50,000 years) for instance and allows to study
geochronology since the formation of the Earth (about 4.5 billion years) until present time.

Historically Zeller~\cite{Zeller} suggested for the first time in 1968 the use of ESR for 
dating geological materials. In 1975 Ikeya~\cite{Ikeya} was able to successfully date stalactite  
in Japanese caves with ESR and in 1978 Robins~\cite{Robins} was even capable of identifying 
ancient heat treatment on flint artefacts with ESR.

More recently, Bourbin \etal~\cite{Bourbin} and Gourier \etal~\cite{Bonduelle} introduced 
a statistical approach based on  estimating an average area  separating any given lineshape 
and the limiting Lorentz (or Breit-Wigner). This measure is a statistical correlation factor 
called $R_{10}$ that we show has several drawbacks and limitations. In sharp contrast, Information Theory (IT) 
provides distances called divergence measures that are able of tackling most of the cases a simple
$R_{10}$ statistical correlation factor is unable to approach. Moreover IT provides accurate bounds to
age from the evaluation of distances with respect to the limiting Lorentz and Gauss distributions.

This work is organized as follows: After describing ESR spectroscopy and spin interactions
classified as dipolar or exchange, we discuss the ESR lineshapes arising in both
situations then move on to the dating procedure based on the evaluation of a statistical correlation
factor $R_{10}$. The latter is unable to describe several important cases (quantum exchange
spin interaction or Gaussian pdf..). Thus we move on to introducing powerful IT tools based on
evaluating distances (also called divergence measures or relative information entropy) between 
different existing pdf's. The latter originate from the ESR lineshape 
by integrating it with respect to the magnetic field. Afterwards we compare these IT results 
to $R_{10}$ values when available. We do not address dosimetry procedures since this is 
beyond the goal of this work assuming that dosimetry has been tackled properly by other works.

\section*{Free Induction Decay function as a Stretched exponential}

Free Induction Decay function $S(t)$ is a spin-spin correlation function 
yielding time-dependent interactions between magnetic moments carried 
by electronic spins in a material.

Since the ESR lineshape is the magnetic field derivative of $S(t)$ Fourier Transform (FT)
(as explained in Supplementary Material) it is capable of of revealing these correlations.

In general, a spin system is expected to interact in two distinct ways:

\begin{enumerate}
\item When electrons are far apart i.e. $r \gg $ Angstr\"om (i.e. several 
hundred Angstr\"oms, microns and beyond) they can be considered
as magnetic moments interacting in a magnetostatic fashion as $\frac{1}{r^3}$. 
This dipolar interaction~\cite{VanVleck} is considered as classical.
\item Quantum exchange~\cite{Bencini} if spins are very close (typical nearest neighbor 
distance about a few Angstr\"oms). When electrons are close, overlap of their wavefunctions 
leads to short-range Slater interaction of the form $\exp(-r/r_0)$ with $r$ the average 
inter-spin distance and $r_0$ on the order of an Angstr\"om.
\end{enumerate}

Actually there is a third mixed regime called DE (Dipolar-Exchange) 
when inter-spin distances are intermediary between the Angstr\"om and large
distances as in the classical regime. This regime~\cite{Kalinikos} which is beyond our scope
is complex since it is a mixture of classical and quantum types.

\begin{figure}[htbp]
\begin{center}
\includegraphics[angle=0,width=80mm,clip=]{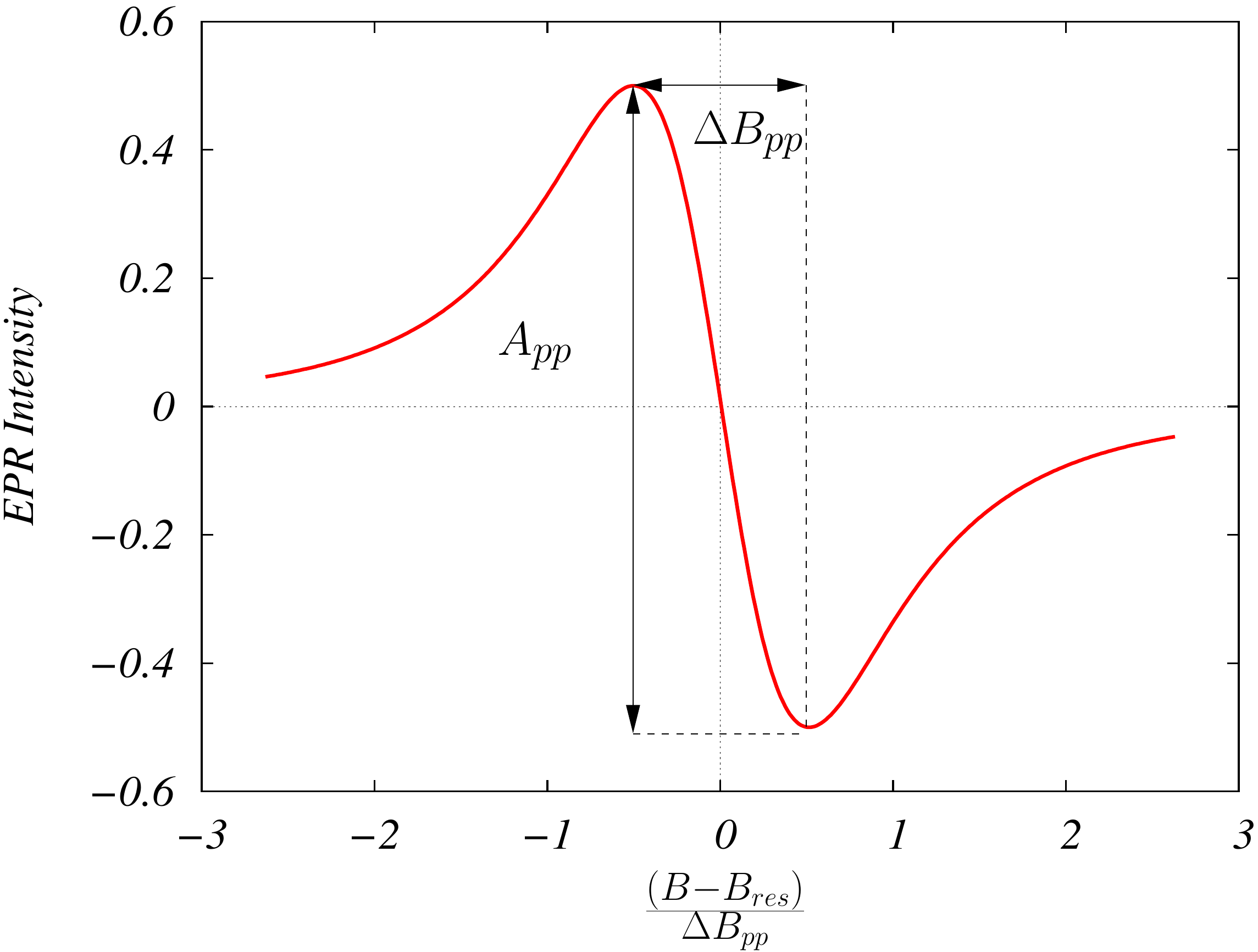}
\end{center}
  \caption{(Color on-line) ESR lineshape key parameters: $B_{res}$ resonance field, 
$A_{pp}$ peak-to-peak amplitude and corresponding linewidth $\Delta B_{pp}$.}
\label{lineshape}
\end{figure}

In ESR or other resonance methods such as ferromagnetic (FMR) or nuclear magnetic 
resonance (NMR), there are, in general, two spin relaxation
times: longitudinal  $T_1$ (along applied external magnetic field) and transverse $T_2$ 
(perpendicular to applied external magnetic field).  \\
 
In the dipolar~\cite{VanVleck} case, Fel'dman and Lacelle~\cite{Feldman} have shown 
that spin correlation function is given by:

\be
S(t)=S(0)\exp[(-t/T_2)^{\beta(D)}]
\ee 

with $T_2$ the transverse relaxation time, a measure of spin density in the sample and 
$\beta(D)= \frac{D}{3}$ with $D$ the dimension of geometrical spin arrangement (3 for 
full spatial, 2 for layers or thin films and 1 for spin chains). \\

In the $D=3$ dipolar case, $\beta= 1$ and $S(t)=S(0)\exp[(-t/T_2)]$
whose FT is a Lorentzian $\hat{S}(\Delta \omega)=\frac{1}{\pi} \frac{T_2}{[1+(\Delta \omega T_2)^2)]} $ 
where $\Delta \omega$ is frequency difference with respect to resonance frequency. \\

In ESR spectroscopy,  the Lorentzian and the Gaussian absorption curves are considered as 
limiting absorption curves whose derivatives with respect to the applied 
field yields the ESR lineshape. 

A Lorentzian appears when we have homogeneous broadening
whereas a Gaussian results (in a solid) from thermal fluctuations of atomic/ionic 
constituents causing changes in the local magnetic field (cf Jonas review~\cite{Jonas}).

\begin{figure}[htbp]
\begin{center}
\includegraphics[angle=0,width=100mm,clip=]{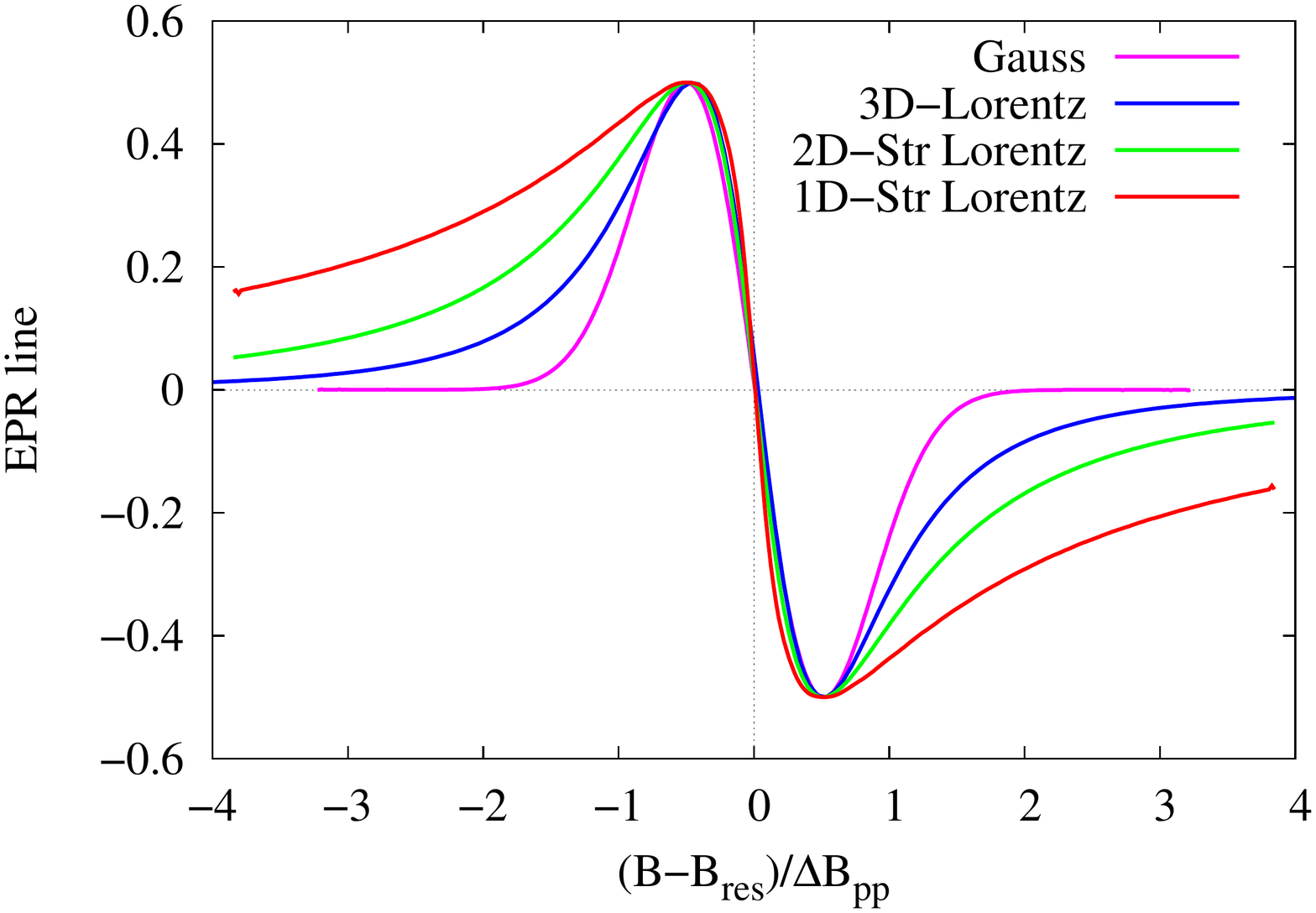}
\end{center}
  \caption{(Color on-line) Illustration of dipolar broadening with ESR lineshapes 
corresponding to spin correlation function
$S(t)=S(0)\exp[(-t/T_2)^{\frac{D}{3}}]$ for a fixed value of parameter $A_{pp}$. The Gaussian  profile is 
the narrowest then we progress toward the broader Lorentzian ($D=3$), stretched  Lorentzian ($D=2$) and finally 
the ($D=1$) stretched  Lorentzian.  {\bf Note}: Str means Stretched.}
\label{dipolar}
\end{figure}

At lower dimensionality $D=1,2$ the rational exponent
$\frac{D}{3}$ leads to a "stretched exponential", "stretched Gaussian" or even
"stretched Lorentzian" (cf fig~\ref{dipolar}) dependence 
\footnote{One might be tempted to believe that since a Lorentzian is given by
$\frac{1}{\pi} \frac{1}{[1+x^2]}$, then a stretched  Lorentzian would be
$\frac{1}{[1+x^\alpha]}$ with $\alpha~\ne~2$. In fact, a
stretched Lorentzian curve has several meanings: one might have a superposition
of Lorentzians or more complicated functions (possibly containing terms such as $\frac{1}{[1+x^\alpha]}$).
Generally, any function whose width is larger than the Lorentzian is considered
stretched.}.\\

Thus "stretched" refers to a spatial arrangement of spins in planes or layers 
($D=2$) or along linear chains ($D=1$).\\

To summarize, $\beta=1$ yields a pure exponential with Lorentzian FT, whereas when
$\beta=2$ we obtain a Gaussian with a Gaussian FT and for intermediate values $1 < \beta <2$,
we recover the stretched curve varieties.\\

Performing the FT (as detailed in Supplementary Material)
using $S(t)=S(0)\exp[(-t/T_2)^{\frac{D}{3}}]$ and
$D=1,2,3,6$ to recover the stretched Lorentzian  ($D=1,2$), 
the Lorentzian ($D=3$) and finally the Gaussian  ($D=6$). 
In addition, one has to determine from the lineshape 
$A_{pp}$ and $\Delta B_{pp}$ that depend on $D, T_2, a$  (cf. fig.~\ref{lineshape}).\\

\begin{figure}[htbp]
\centering
\includegraphics[angle=0,width=100mm,clip=]{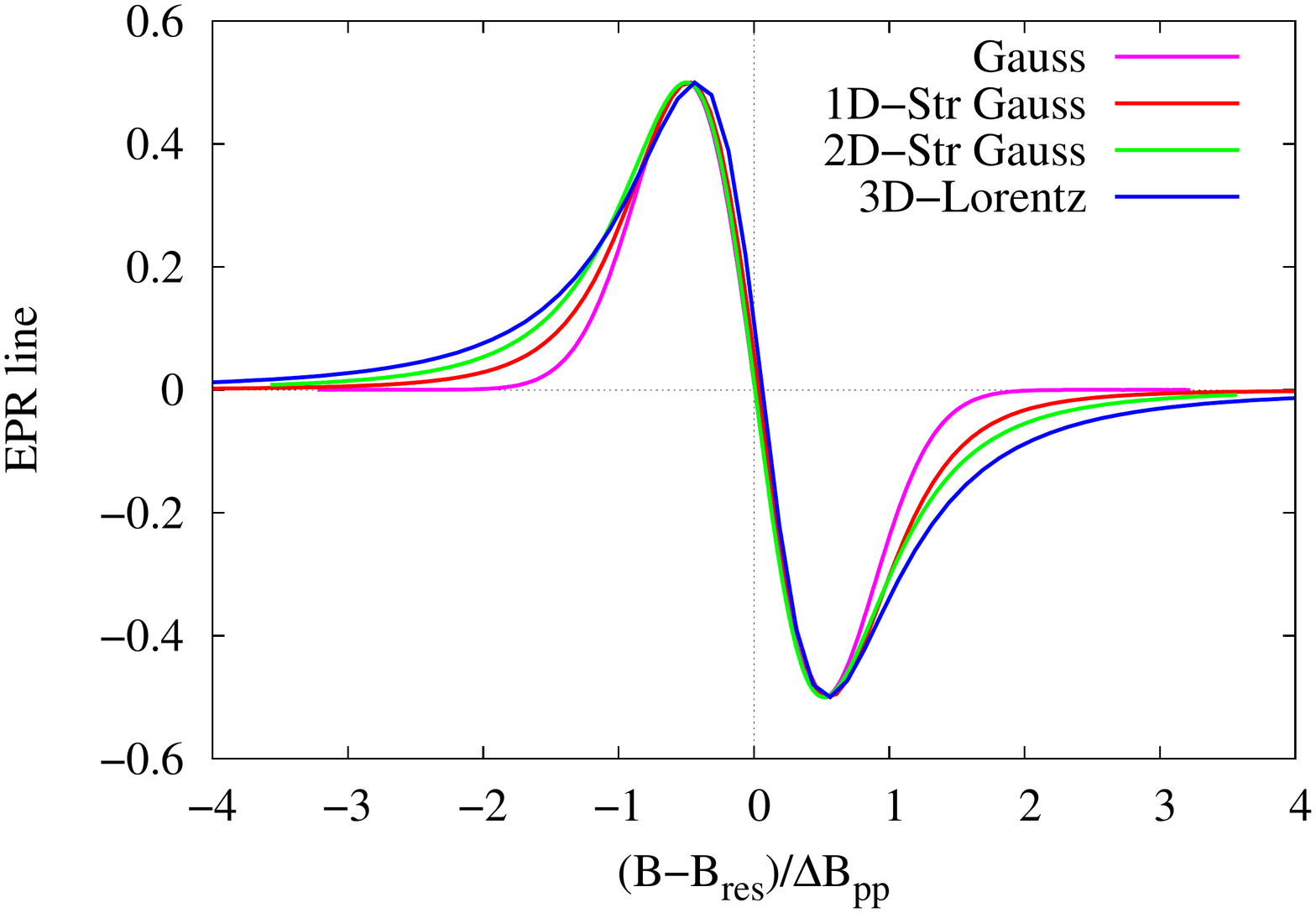}
  \caption{(Color on-line) Illustration of exchange narrowing with ESR lineshapes derived from 
$S(t)=S(0)\exp[(-t/T_2)^{\beta^*(D)}]$ with $\beta^*(D)$ function of dimension $D$. 
Lineshapes evolve to broader (Stretched)  from narrowest (Gaussian) as dimensionality 
increases ($D=1,2,3$) in sharp contrast with respect to the dipolar case (fig.~\ref{dipolar}). 
When $D=3$ the Lorentzian is recovered as in the dipolar case.}
\label{exchange}
\end{figure}

ESR lineshape results displayed in fig.~\ref{dipolar} show that we evolve from
Gaussian to Lorentzian  ($D=3$), to stretched  Lorentzian ($D=2$) and finally
stretched  Lorentzian ($D=1$). Comparing to the quantum or "spin diffusion" case (fig.~\ref{exchange}) 
based on spin correlation function $S(t)=S(0)\exp[(-t/T_2)^{\beta^*(D)}]$ with $\beta^*(D)$ 
function of spin arrangement dimension $D$. $\beta^*(1)=3/2$ in 1D (see Dietz \etal~\cite{Dietz}) 
and $\beta^*(2)=1$ in 2D (a simplified version of the very complex 2D case (see Richards \etal~\cite{Richards}). \\

Classical dipolar interactions lead to spectrum broadening~\cite{Kittel}
i.e. damping. In sharp contrast, quantum exchange among spins tends to reduce the ESR linewidth
("Exchange Narrowing" effect) and consequently, damping.

\section*{Dating with $R_{10}$ a statistical correlation measure}
 
Characterizing the ESR lineshape around resonance,
Bourbin \etal~\cite{Bourbin} and Gourier \etal~\cite{Bonduelle} introduced a statistical
method based on an average area estimation between any given lineshape and the 3D Lorentzian.\\

They introduced after ref.~\cite{Dietz,Richards} functional transformations on $B$ and 
function $F(B)$, the derivative with respect to magnetic field of the 
$S(t)$) FT in order to determine age from ESR lineshape from some universal behavior.

The double scaling transformation (cf. Supplementary Material) results in a scaled
magnetic field $B \rightarrow x$ and a scaled function $F(B)\rightarrow f(x)$ 
such that "Lorentzian derivative" $F(B)$ corresponds to linear function $f_L(x)=x+3/4$ 
whereas  "Gaussian derivative"  corresponds to $f_G(x)=\exp(x-1/4)$. Then a correlation  
factor $R_{10}$ is extracted from:

\be
R_{10}= \frac{1}{10} \int_{0}^{10} [f(x)-f_L(x)] dx
\ee

meaning that the area estimation is some kind of distance measure separating the lineshape from the
Lorentzian. The value 10 is the largest statistically estimated $x$ value.

When $R_{10} > 0$ we have close-neighbor spins regime (quantum exchange interaction) 
with a mixed Gaussian-Lorentzian profile.

Finally, when $R_{10} < 0$ we have distant spins (dipolar regime) in lower dimension
($D=1,2$) with a stretched Lorentzian  lineshape.

\begin{figure}[htbp]
\centering
\includegraphics[angle=0,width=100mm,clip=]{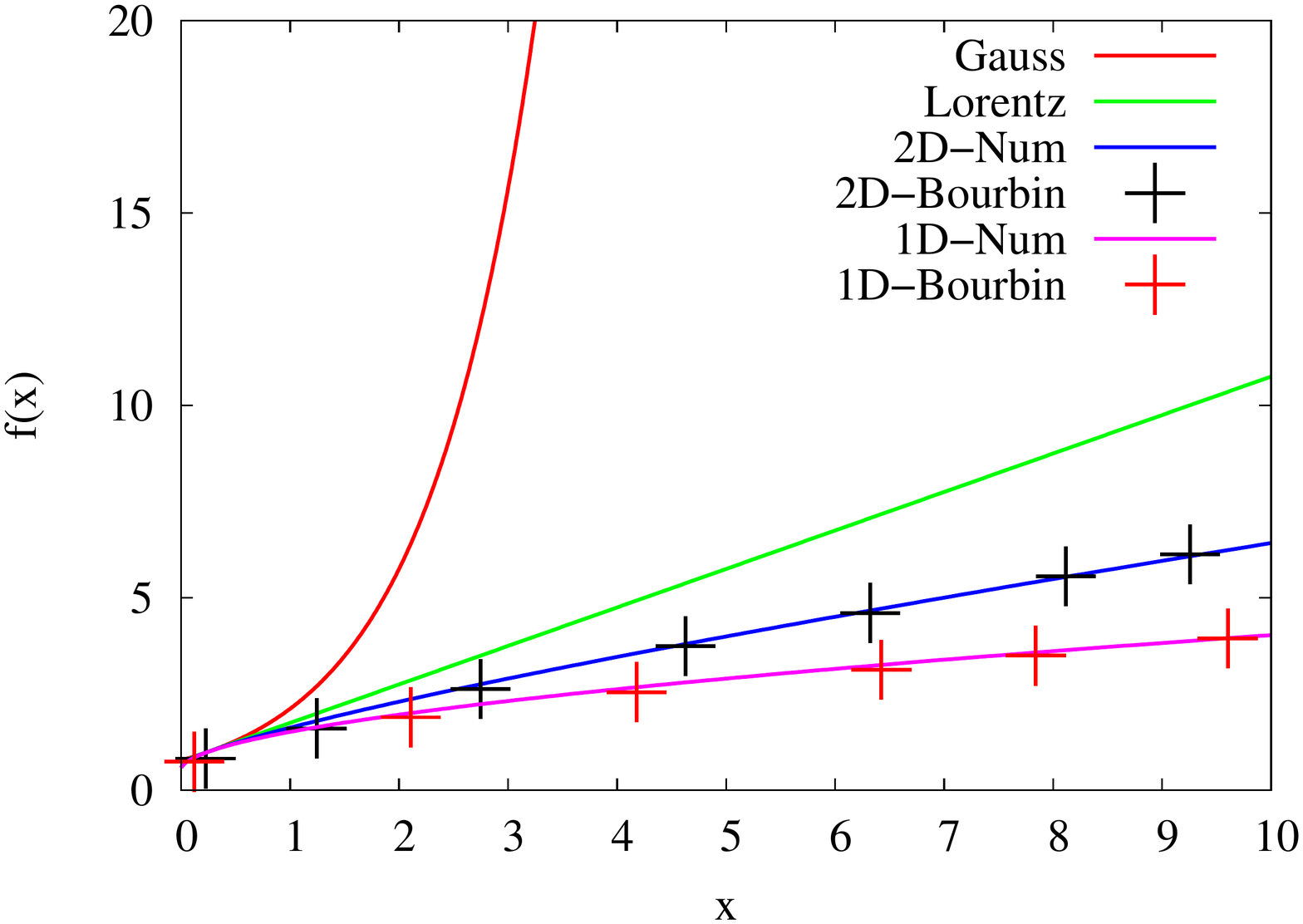}
 \caption{(Color on-line) $f(x)$ functions corresponding to the different $F(B)$.
"Lorentzian derivative" corresponds to $f_L(x)=x+3/4$ whereas " Gaussian derivative" corresponds to 
$f_G(x)=\exp(x-1/4)$. Our results agree with Bourbin \etal~\cite{Bourbin} numerical calculations.}
\label{diagdip}
\end{figure} 

\begin{figure}[htbp]
\centering
\includegraphics[angle=0,width=100mm,clip=]{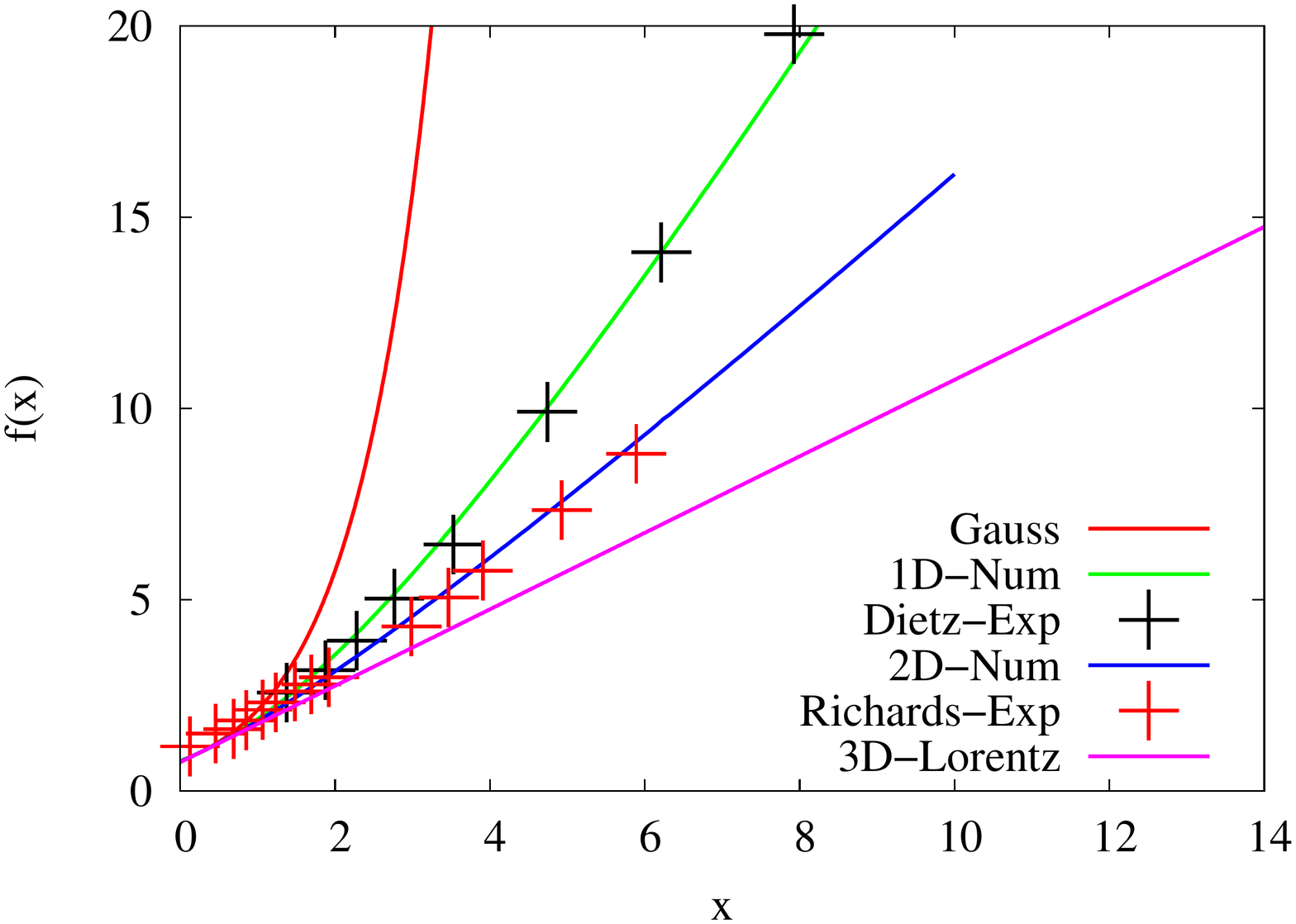}
\caption{(Color on-line) In the quantum case $D=1,2$ curves  lie between the
Lorentzian and the Gaussian and our results agree with
experimental data of Dietz \etal~\cite{Dietz} in 1D for (CH$_3$)$_4$NMnCl$_3$ as well as with
Richards \etal~\cite{Richards} in 2D for K$_2$MnF$_4$. {\bf Note}: Num stands for numerical and Exp for experimental.}
\label{diagex}
\end{figure}

In the dipolar case, $x,f(x)$ diagrams (cf fig.~\ref{diagdip}) serve to identify $R_{10} > 0$ region
between Gaussian and Lorentzian whereas  $R_{10} < 0$ is below the Lorentzian.

In sharp contrast, $x, f(x)$ diagrams in the exchange case for $D=1,2$ 
(region $R_{10} > 0$) are  between the Lorentzian and the Gaussian (cf fig.~\ref{diagex}). Thus
one has to distinguish the classical dipolar from the quantum exchange case when encountering $R_{10} > 0$.

Nevertheless, obtaining $R_{10}$ from ESR lineshape is interesting since it allows determination of the
nature of spin interactions as quantum or classical (from $S(t)$) and their geometrical arrangement
from the $D$ value as done by Bourbin \etal~\cite{Bourbin} and Gourier \etal~\cite{Bonduelle} 
who considered only the classical dipolar case $D=1,2$, thus the need to examine the quantum picture 
(cf fig.~\ref{exchange}).

In the intermediate dipolar case (lower dimensionality) $D=1,2$ lines are under the Lorentzian,
whereas in the intermediate exchange~\cite{Dietz,Richards} case $D=1,2$ lines are between the 
Gaussian and the Lorentzian.

This is important in order to obtain a reliable and precise dating assessment of rocks 
and sediments along the 
lines of Bourbin \etal~\cite{Bourbin} and Gourier \etal~\cite{Bonduelle} who derived two different sets
of age formulae from the $R_{10}$ factor.

\section*{Information theory distance measures}

Introducing a measure based on Information Theory~\cite{Cover} helps to discriminate between the 
classical and quantum pictures.

There are several divergence or distance measures between probability densities such as the
Kullback-Leibler~\cite{Press,Cover}, squared Hellinger, $\alpha$-divergence, 
Jensen-Shannon, total variation~\cite{Cover,Morales}...(see Supplementary Material).

We choose the Cauchy-Schwarz divergence (CSD) measure~\cite{Press,Cover} defined between two 
probability distributions $P,Q$ as:
\bea
D(P||Q)=-\ln\left(\int_{x \in {\mathcal X}} dx \; p(x)q(x)\right)+ \nonumber  \\
\frac{1}{2}\ln\left(\int_{x \in {\mathcal X}}dx \; p^2(x)\right)+
\frac{1}{2}\ln\left(\int_{x \in {\mathcal X}}dx \; q^2(x)\right)
\eea

where ${\mathcal X}$ is the set of values taken by  
continuous probability densities (pdf) $p(x),q(x),~x~\in~{\mathcal X}$.

CSD obeys several axioms of distance (see Supplementary Material) such as positivity 
$D(P||Q)\ge 0$, and triangle inequality: $D(P||R) \le D(P||Q)+ D(Q||R)$. 
It obeys also symmetry  $D(P||Q)=D(Q||P)$ unlike 
the Kullback-Leibler (KL) measure~\cite{Press,Cover} defined by:
\be
D_{KL}(P||Q)=\int_{x \in {\mathcal X}} dx \; p(x) \ln \frac{p(x)}{q(x)}
\ee
Moreover CSD does not suffer from singularities whereas KL does.

The pdf are obtained from the ESR lineshape by integrating it with respect to the
magnetic field (Detailed procedure is described in Supplementary Material).

In Table~\ref{Lorentz CSD} we display CSD distances between a unit Lorentzian pdf and a set
of Dipolar and Exchange pdf along with the $R_{10}$ factor and corresponding ages.

In Table~\ref{Gauss CSD} we display CSD distances between a unit Gaussian and a set
of Dipolar and Exchange pdf along with the $R^*_{10}$ factor and corresponding ages
that would correspond to formula:

\be
R^*_{10}= \frac{1}{10} \int_{0}^{10} [f(x)-f_G(x)] dx, \, f_G(x)=\exp(x-1/4)
\ee

Actually, we did not use the above formula and rather extrapolated the Lorentzian $R_{10}$ 
values to estimate the age by modifying Bourbin~\etal \,coefficient $\alpha_B$. In addition 
we provide an experimental method (in Supplementary Material) 
to determine age and evaluate its bounds based on (CSD) distances 
$d_L$  and  $d_G$ with respect to unit Lorentzian and Gaussian
distributions labeled as $p(x)$ as well as from $\sigma_q$ the standard deviation of 
$q(x)$ corresponding to the measured  ESR spectrum. 
Finally, age is expressed with the formula $10^{A_{L,G}}$ 
where $A_{L,G}=d_{L,G}^\xi \sigma_q^\eta$ with exponents $(\xi,\eta)$ determined by optimization 
(procedure detailed in Supplementary Material).

\begin{table}[htbp]
\begin{center}
\begin{tabular}{| l| c| c| c|}
\hline
Lorentzian $P$ & $D(P||Q)$  & $R_{10}$  & Age (Gyr)\\
\hline
\hline
Dipolar pdf $Q$  &  &  & \\
\hline
\hline
$D=1$  Str Lorentz & 0.21 &  -2.94  &  3.54\\
$D=2$  Str  Lorentz& 2.50 $\times 10^{-2}$ & -1.94 & 2.74 \\
$D=3$  Lorentz& 1.19 $\times 10^{-6}$   & 0.0  & 1.67\\
Gauss&  2.55 $\times 10^{-2}$  &  1.71 $\times 10^{3}$ & Undefined\\
\hline
\hline
Exchange pdf $Q$ &  &  & \\
\hline
\hline
$D=1$  Str Gauss&  1.14 $\times 10^{-2}$  & 6.78 & 0.29 \\
$D=2$  Str Gauss &   4.12 $\times 10^{-3}$ & 2.30 & 0.93\\
$D=3$  Lorentz&  1.19 $\times 10^{-6}$  & 0.0 & 1.67 \\
Gauss&  2.55 $\times 10^{-2}$  &  1.71 $\times 10^{3}$ & Undefined\\
\hline
\end{tabular}
\caption{\label{Lorentz CSD} Cauchy-Schwarz distances separating a unit width $P$ Lorentzian and $Q$ distributions (Dipolar and Exchange) with corresponding $R_{10}$ factor yielding age in Gyr (Giga or billion years). 
Age in years is deduced from Bourbin~\etal~\cite{Bourbin} formula $10^{\left(\frac{R_{10}-\beta_B}{\alpha_B}\right)}$ with $\alpha_B=-9.$ and $\beta_B=83.$. Note that 
Skrzypczak~\etal~\cite{Bonduelle} define another set of coefficients: $\alpha_S=-5.3$ and $\beta_S=48.9$ (Other examples are detailed in Supplementary Material).} 
\end{center}
\end{table}

\begin{table}[htbp]
\begin{center}
\begin{tabular}{| l| c| c| c|}
\hline
Gaussian $P$ & $D(P||Q)$ & $R^*_{10}$  & Age (Myr) \\
\hline
\hline
Dipolar pdf $Q$ &   & &     \\
\hline
\hline
$D=1$  Str Lorentz &  0.25  & -2.94 & 90.59 \\
$D=2$  Str  Lorentz&   5.11 $\times 10^{-2}$ & -2.89 & 89.61 \\
$D=3$  Lorentz&   1.51 $\times 10^{-2}$ & 2.73 & 27.09 \\
Gauss&   0.0  & 0.0  & 48.44 \\
\hline
\hline
Exchange pdf $Q$ &    & &   \\
\hline
\hline
$D=1$  Str Gauss&  1.86 $\times 10^{-2}$ & -5.02 $\times 10^{-2}$ & 48.96    \\
$D=2$  Str Gauss &  1.44 $\times 10^{-2}$   & 3.51 & 22.91  \\
$D=3$  Lorentz &  1.51 $\times 10^{-2}$  & 2.73 & 27.09   \\
Gauss&   0.0 & 0.0  & 48.44 \\
\hline
\end{tabular} 
\caption{\label{Gauss CSD} Cauchy-Schwarz distances separating a unit standard deviation $P$ Gaussian 
and $Q$ distributions (Dipolar and Exchange). The corresponding $R^*_{10}$ factor is obtained from extrapolating the above Lorentzian $R_{10}$ value and the age obtained in Myr (Mega or million years)
is deduced from altering the Bourbin~\etal \,coefficient $\alpha_B$ (Other examples are detailed in Supplementary Material).}
\end{center}
\end{table}

In this work we developed an ESR lineshape based chronometric dating methodology handling
both Dipolar (classical) and Exchange (quantum) interactions between unpaired spins.

It is based on evaluating IT distances (divergence measures) 
between a Lorentzian or a Gaussian (analytical) pdf's and an experimental pdf obtained from the 
measured ESR lineshape integrated over the applied magnetic field. 

From the ESR detection point of view, ancient material age is extracted with FT of
a spin-correlation function $S(t)$ pertaining to unpaired interacting spins that 
might have extremely long lifetimes and originate from ($\alpha, \beta$ and $\gamma$) 
ionizing radiation.

$S(t)$ is of a stretched exponential form with a characteristic exponent $\beta(D)$ 
that depends on unpaired spin arrangement dimension $D$. 

In the dipolar interaction case, $\beta(D)=\frac{D}{3}$ according to
Fel'dman~\etal~\cite{Feldman}  whereas in the
exchange interaction case, Dietz~\etal~\cite{Dietz} and Richards~\etal~\cite{Richards}
found $\beta^*(D=1)=3/2$ and $\beta^*(D=2)=1$ respectively (in the "spin diffusion" case).

Considering aging as an evolutionary IT distance (or relative information
entropy) between pdf's stemming from ESR 
lineshapes integrated with respect to the applied magnetic field brings a major
paradigm shift to chronometric dating.

IT provides an alternate look at dating based on using distances (with respect to
Gaussian, Lorentzian...) as well as arbitrary coupling between spins (dipolar, exchange...)
paving the way to cover any period of time with the proviso of having performed 
previously an appropriate dosimetry analysis.  IT approach might be further extended to the
intermediary Dipolar-Exchange regime where classical and quantum interactions are 
intertwined.\\

\section*{Acknowledgments}
We would like to thank  Prof. Herv\'e Bellon and Dr. St\'efan Lalonde for suggesting this problem
and many enlightening discussions.


Materials and Methods\\
Information theory effective distance measure\\
Experimental protocols for dating with IT distance\\
Figs. S1 to S4\\
Tables S1 to S2\\
References  \textit{(19-26)} and \textit{(29-35)}

\nocite{*}
\bibliographystyle{Science}
\bibliography{paper}

\section*{Supplementary Materials}

\setcounter{equation}{0}
\setcounter{figure}{0}
\setcounter{table}{0}
\setcounter{page}{1}
\setcounter{section}{0}
\makeatletter
\renewcommand{\thesection}{S\arabic{section}}
\renewcommand{\theequation}{S\arabic{equation}}
\renewcommand{\thefigure}{S\arabic{figure}}
\renewcommand{\thetable}{S\arabic{table}}

\newcommand{\bibnumfmt}[1]{[S#1]}
\newcommand{\citenumfont}[1]{S#1}

\section*{Materials and Methods}

ESR absorption spectroscopy reveals time-dependent interactions between electronic spins
carrying magnetic moments in a material as reflected in the correlation function
$S(t)$ also called Free Induction Decay function.

\begin{figure}[htbp]
\begin{center}
\includegraphics[angle=0,width=100mm,clip=]{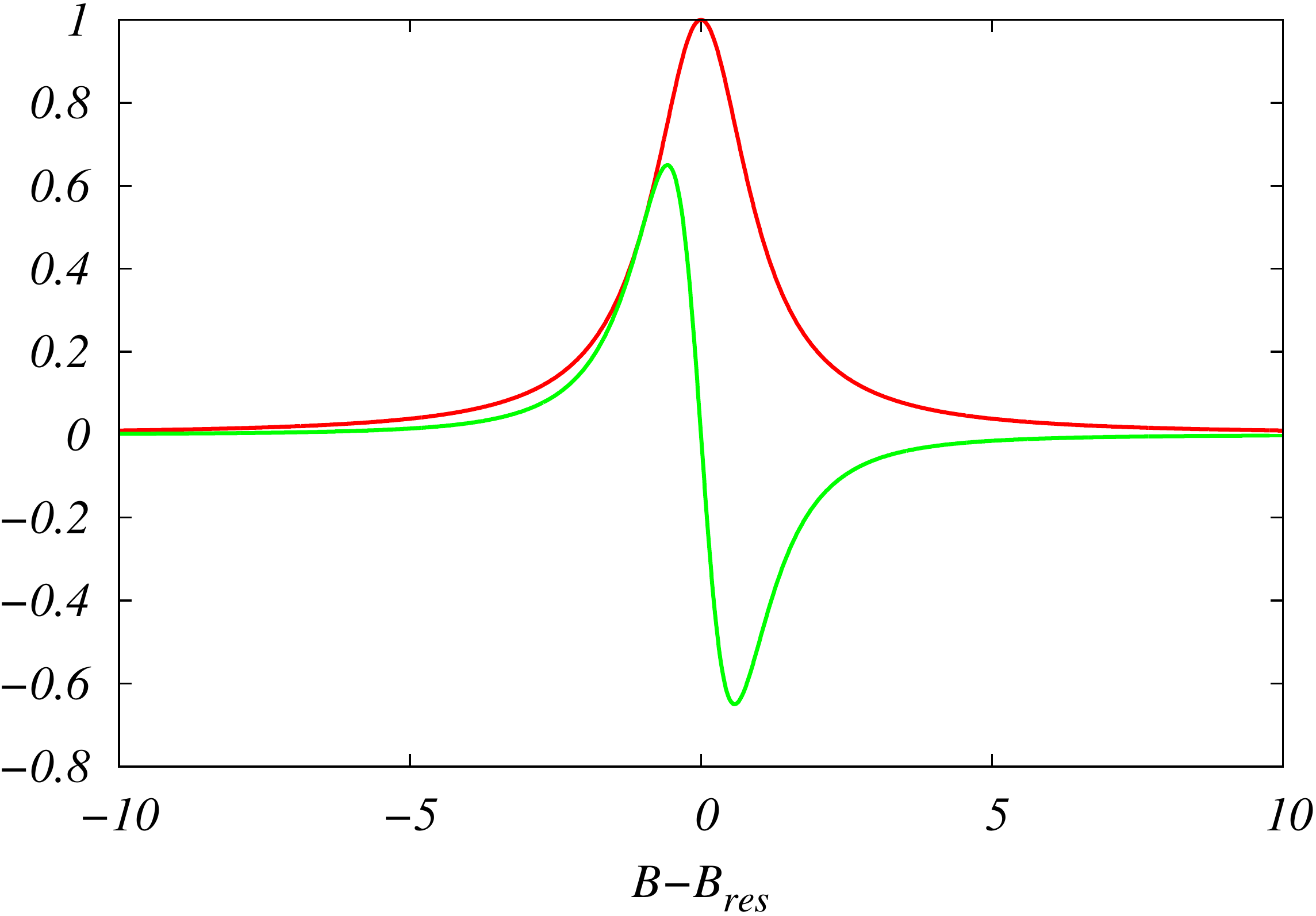}
\end{center}
  \caption{(Color on-line) absorption line (in red) and its field derivative the ESR 
lineshape (in green) in the Lorentzian case.}
\label{lorentz}
\end{figure}

The ESR lineshape is the field derivative of $S(t)$ Fourier Transform (FT). An example
is shown in the case of the Lorentzian in fig.~\ref{lorentz}.

Magnetic resonance at a field $B_{res}$ is such that frequency $\omega=\gamma B$ where $\gamma$ 
is gyromagnetic ratio
given by $\gamma=\mu_0 g_L e \hbar/2 m_e$ ($\mu_0$= vacuum permeability, $g_L=$ Land\'e factor, 
$e,m_e=$ charge and electron mass).  This implies  $\Delta \omega \propto  B-B_{res}$
with  $B$ (applied field) and resonance field $B_{res}$.\\

When $S(t)=S(0)\exp[(-t/T_2)]$ its FT is a Lorentzian in $B$ such that: 
$\hat{S}(B)=\frac{1}{\pi} \frac{1}{\left[1+\left(\frac{B-B_{res}}{\Delta B_{pp}}\right)^2\right]} $.  \\

In the general case, ESR lineshape is obtained by taking the derivative with respect to applied field $B$) of 
$\hat{S}(B)$ such that $F(B)=\frac{d\hat{S}(B)}{dB}$. Note that $\hat{S}(B)$ is often denoted as
the ESR power absorption, the imaginary part of the generalized susceptibility $\chi"(B)$.\\

$S(t)$ FT is written as $\hat{S}(\omega)=\int_{0}^{+\infty}e^{-i\omega t} S(t) dt$.

In ESR, $\omega$ is proportional to the applied magnetic field allowing us to replace $\omega$
by $B-B_{res}$ resulting in $\hat{S}(B-B_{res})=\int_{0}^{+\infty}e^{-ia(B-B_{res}) t} S(t) dt$
with $a=g_L\mu_B/\hbar$, $g_L$  the Land\'e factor, $\mu_B$ Bohr magneton and $\hbar$ Planck constant. \\

ESR lineshape $F(B)$ is obtained from the derivative with respect to $B$ of $\hat{S}(B-B_{res})$ yielding:

\bea
F(B)&=&Re \left[\int_{0}^{+\infty}\frac{d}{dB}e^{-ia(B-B_{res}) t} S(t) dt \right] \nonumber \\
   &=&  -a \int_{0}^{+\infty} t \sin[ a (B-B_{res}) t] S(t) dt
\eea

Hence it suffices to evaluate the above integral using $S(t)=S(0)\exp[(-t/T_2)^{\frac{D}{3}}]$ and
$D=1,2,3,6$ to recover the stretched Lorentzian  ($D=1,2$), 
the Lorentzian ($D=3$) and finally the Gaussian  ($D=6$). \\

Integrating requires special methods to reduce errors resulting from large number of oscillations
emanating from the term $\sin[ a (B-B_{res}) t]$. We have used a number of numerical 
algorithms~\cite{Press} in order to obtain accurate and reliable results~\cite{Helfand,Montroll,Wuttke}.

In order to characterize the ESR lineshape around resonance,
Bourbin \etal~\cite{Bourbin} and Gourier \etal~\cite{Bonduelle} introduced a statistical
method based on an average area estimation between any given lineshape and the 3D Lorentzian.\\

They introduced after ref.~\cite{Dietz,Richards} a double scaling transformation acting on the
magnetic field $B$ and the ESR lineshape $F(B)$ such that:

\bea
x&=&\left(\frac{B-B_{res}}{\Delta B_{pp}}\right)^2,  \nonumber \\
y&=&f(x)=\sqrt{\frac{A_{pp}}{|F(B-B_{res})|}\frac{|B-B_{res}|}{\Delta B_{pp}} }
\eea

For example,  "Lorentzian derivative" $F(B)$ corresponds to linear function $f_L(x)=x+3/4$ 
whereas  "Gaussian derivative"  corresponds to $f_G(x)=\exp(x-1/4)$.

Dating is based on a statistical correlation  factor $R_{x_{m}}$ extracted from the integral:

\be
R_{x_{m}}= \frac{1}{x_{m}} \int_{0}^{x_{m}} [f(x)-f_L(x)] dx
\ee

where $x_{m}$ is some assumed large $x$ value such that the area estimated lying between
$f(x)$ and $f_L(x)]$ is a distance measure separating the ESR lineshape from the
Lorentzian. 

Note that it is possible to determine another statistical correlation  factor $R^*_{x_{m}}$ 
extracted from the integral:

\be
R^*_{x_{m}}= \frac{1}{x_{m}} \int_{0}^{x_{m}} [f(x)-f_G(x)] dx
\ee

where the area value indicates how much the ESR lineshape differs from the Gaussian. 
 

\subsection*{Information theory effective distance measure}
Information Theory (IT) provides an effective distance~\cite{Cover} or a divergence measure
separating two functions (in this case probability density functions or pdf) to
enable their comparison not only from the geometrical shape point of view but
also from the information content as well. 

Going from a distance connecting two points to one linking two functions helps
determine how much the functions are dissimilar in several ways
(analytical, geometrical and informational). That is why $D(P||Q)$
is also called {\it relative information entropy} that does not
behave as a metric but rather as the square of the Euclidian one~\cite{Cover}. 

Several examples of $D(P||Q)$ are displayed in Table~\ref{Divergences}.

\begin{table}[htbp]
\begin{center}
\begin{tabular}{| l| c|}
\hline
Distance $D(P||Q)$         &           Definition    \\ 
\hline                    
Squared Hellinger & $  \int_{x \in {\mathcal X}} dx \; (\sqrt{ p(x)} - \sqrt{q(x)})^2   $ \\                      
Squared triangular  &  $ \int_{x \in {\mathcal X}} dx \; \frac{ (q(x)-p(x))^2  }{ p(x)+q(x)}  $ \\  
Squared perimeter   &  $ \int_{x \in {\mathcal X}}dx \; \sqrt{p^2(x)+q^2(x)} -\sqrt{2} $ \\                    
$\alpha$-divergence  &  $ \frac{4}{1-\alpha^2} \left( 1-\int_{x \in {\mathcal X}} dx \; p^\frac{1-\alpha}{2}(x)q^{1+\alpha}(x)\right) $ \\  
Jensen-Shannon  &   $\frac{1}{2} \int_{x \in {\mathcal X}} dx \; \left(p(x) \ln \frac{2p(x)}{ p(x)+q(x)} + q(x) \ln \frac{2q(x)}{ p(x)+q(x)} \right)$ \\   
Pearson $\chi^2_P$  &  $ \int_{x \in {\mathcal X}} dx \; \frac{ (q(x)-p(x))^2  }{ p(x)}  $ \\  
Neyman $\chi^2_N$   &  $ \int_{x \in {\mathcal X}} dx \; \frac{ (q(x)-p(x))^2  }{ q(x)}  $ \\   
Total variation (metric) &   $\frac{1}{2}  \int_{x \in {\mathcal X}} dx \; |p(x) - q(x)|$ \\  
\hline
\end{tabular} 
\caption{\label{Divergences} Examples of Information Theory distances between two pdf $p(x), q(x)$.
Note that $\alpha$-divergence, Pearson $\chi^2_P$ and Neyman $\chi^2_N$ do not satisfy 
the axiomatic symmetry requirement of a distance $D(P||Q)=D(Q||P)$.}
\end{center}
\end{table}

The Cauchy-Schwarz distance (CSD) measure~\cite{Cover,Press} is convenient since
it obeys the axioms of distance and is free of singularities (see main article).

ESR absorption spectra when interpreted as a pdf (see fig.~\ref{lorentz} for 
the Lorentzian case), is considered as modified by aging due to time-dependent 
interactions between electronic spins in an archaeological material. 

The time correlation function $S(t)$ (Free Induction Decay function) yields the  
ESR lineshape $F(B)$ since it is the field derivative of its Fourier Transform (FT).

Thus an IT distance could be used as a measure of aging induced by ionizing radiation
and this supplement details and illustrates the procedure for carrying the 
dating method. 

\subsection*{Experimental protocols for dating with IT distance}
Experimentally, two procedures are possible: 
\begin{enumerate}
\item Direct treatment of the absorption spectra considered as a pdf and evaluating
the IT distance with respect to a given reference (Lorentzian, Gaussian...).
\item Extraction of the absorption spectra from the measured ESR lineshape $F(B)$ by
integrating it with respect to the applied field $B$ and afterwards evaluating
the IT distance.
\end{enumerate} 

The extraction of the absorption spectrum from ESR lineshape $F(B)$ entails several steps:
\begin{enumerate}
\item Spline interpolation of the ESR lineshape $F(B)$.
\item Symmetrization of the ESR lineshape using shifting by $\delta$ such that 
$\int F(B) dB= \delta \int dB$.
\item Field integration of the symmetrized ESR lineshape to get the absorption spectrum.
\item Filtering (smoothing) of the absorption spectrum.
\item Normalization of the absorption spectrum and transformation into a pdf $p(x)$ 
with $x=\left(\frac{B-B_{res}}{\Delta B_{pp}}\right)$.
\end{enumerate}

After transformation of the ESR lineshape into a pdf, the evaluation of the CSD
measure~\cite{Cover,Press} can be undertaken to determine the 
age from the evaluation of the CSD measure taken between the different 
probability distributions~\cite{Helfand,Montroll,Wuttke} corresponding to the ESR spectrum 
and the Lorentzian or Gaussian considered as limiting distributions.

Age is determined from (CSD) distances $d_L$  and  $d_G$ with respect to unit Lorentzian and Gaussian
distributions labeled as $p(x)$ as well as from $\sigma_q$ the standard deviation of 
$q(x)$ corresponding to the measured  ESR spectrum. Age is evaluated with the formula $10^{A_{L,G}}$ 
where $A_{L,G}=d_{L,G}^\xi \sigma_q^\eta$ for the Gyr (Giga or billion year), 
Myr (Mega or million year) cases whereas
it is given by $A_{L,G}=d_{L,G}^\zeta \sigma_q^\eta$ for the Kyr (thousand year) cases with $d=d_L$ in 
the Lorentz or $d=d_G$ in the Gaussian case. Exponents $\xi$=-1.28,  $\eta=2.74 \times 10^{-2}$,  
$\zeta=$-0.32 are determined by optimization (see Table~\ref{Ages}).

\begin{table}[htbp]
\begin{center}
\begin{tabular}{|l| c|c|c| c|c|c|}
\hline
Sample &  $d_L$  &  $d_G$ & $\sigma_q$  & Lorentz Age  & Gauss Age  \\
\hline 
Gunflint~\cite{Skrzypczak} &  0.18   & 0.23  &   8.17  &  2.34     Gyr  &  8.91 $\times 10^{-3}$  Gyr\\
B4~\cite{Balan} &  0.21  &  0.21  & 0.77  &  8.51     Myr &   11.78    Myr  \\
Enamel~\cite{Jonas}    & 1.49 $\times 10^{-2}$  &  1.17 $\times 10^{-2}$  & 1.60  &  7.29    Kyr  & 14.47  Kyr  \\
\hline
\end{tabular} 
\caption{\label{Ages} Age is evaluated with the formula $10^{A_{L,G}}$ where $A_{L,G}=d_{L,G}^\xi \sigma_q^\eta$ 
for the Gyr, Myr cases whereas it is given by $A_{L,G}=d_{L,G}^\zeta \sigma_q^\eta$ for the Kyr cases with $d=d_L$ for the Lorentz case or $d=d_G$ for the Gaussian case.}
\end{center}
\end{table}

We illustrate the Information Theory dating procedure with a first example pertaining to Gunflint from 
Schreiber beach locality (Port Arthur Homocline) in Ontario Canada.

\begin{figure}[htbp]
\begin{center}
\includegraphics[angle=0,width=100mm,clip=]{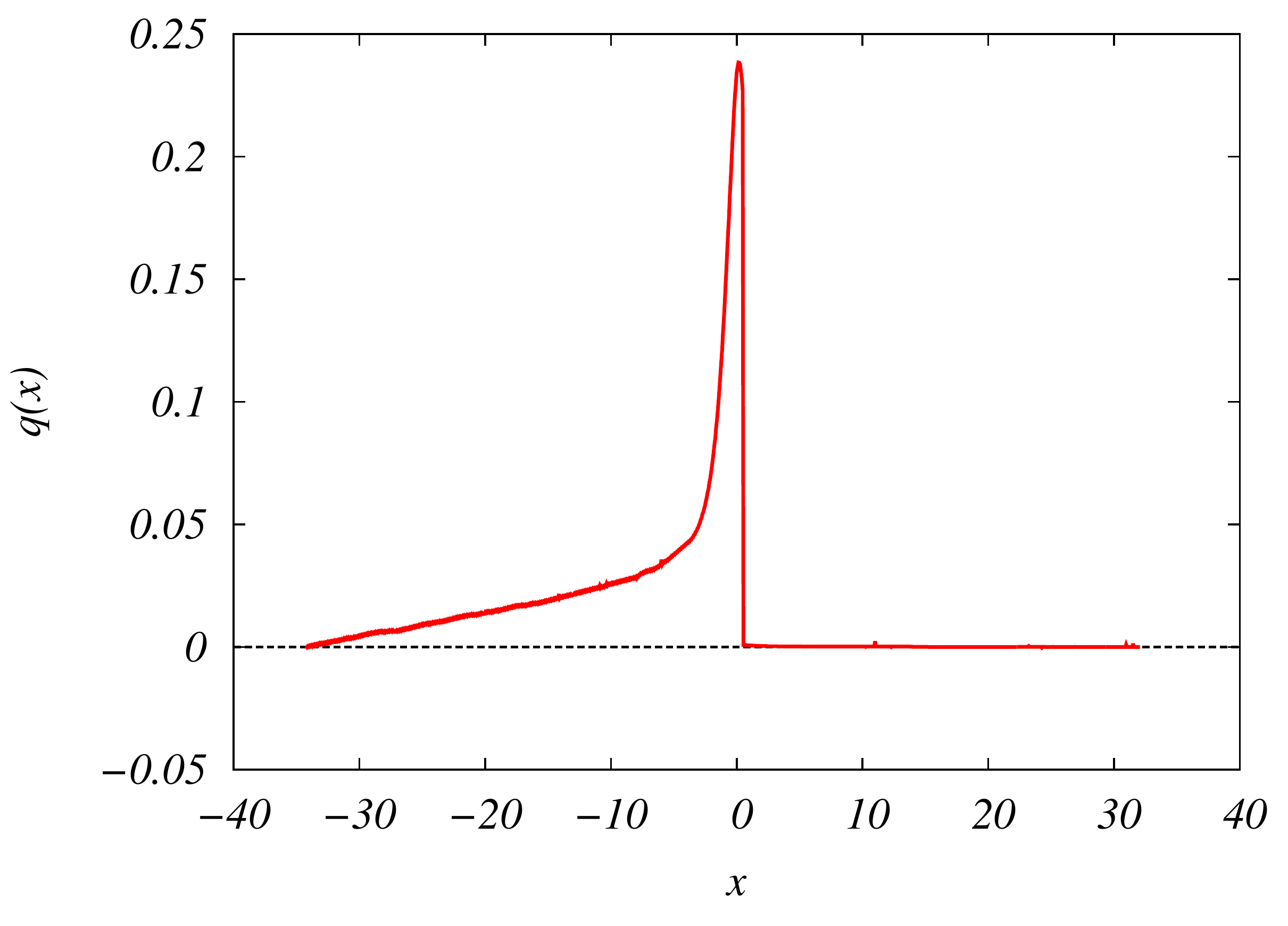}
\end{center}
  \caption{(Color on-line) Absorption line obtained from the ESR lineshape Gunflint
sample taken from Skrzypczak PhD thesis~\cite{Skrzypczak} fig.II.25}
\label{Gunflint}
\end{figure}

After extracting the absorption spectrum from the
ESR lineshape, we find the CSD between Skrzypczak Gunflint~\cite{Skrzypczak} $q(x)$ and a unit 
Lorentzian $p(x)$ as 0.18  whereas it is 0.23 between  $q(x)$  and a unit Gaussian  $p(x)$.

In order to extract the age based on the $R_{10}$ evaluation as
explained in the main work, we find:  $R_{10}$=~-1.765 with age=2.62 Gyr.
Lorentz and Gauss CSD ages are 2.34  Gyr  and  8.91 $\times 10^{-3}$  Gyr respectively 
(cf. Table~\ref{Ages}) whereas  Skrzypczak-Bonduelle estimated it to be about 1.88~Gyr 
(Gyr) in her PhD thesis~\cite{Skrzypczak}.

Another example is a latosol sediment called B4 drawn from Balan \etal~\cite{Balan} fig.2
and extracted from 100 cm depth at a site 60 km North of the city of Manaus capital of 
Amazonas state in North-Western Brazil.

\begin{figure}[htbp]
\begin{center}
\includegraphics[angle=0,width=100mm,clip=]{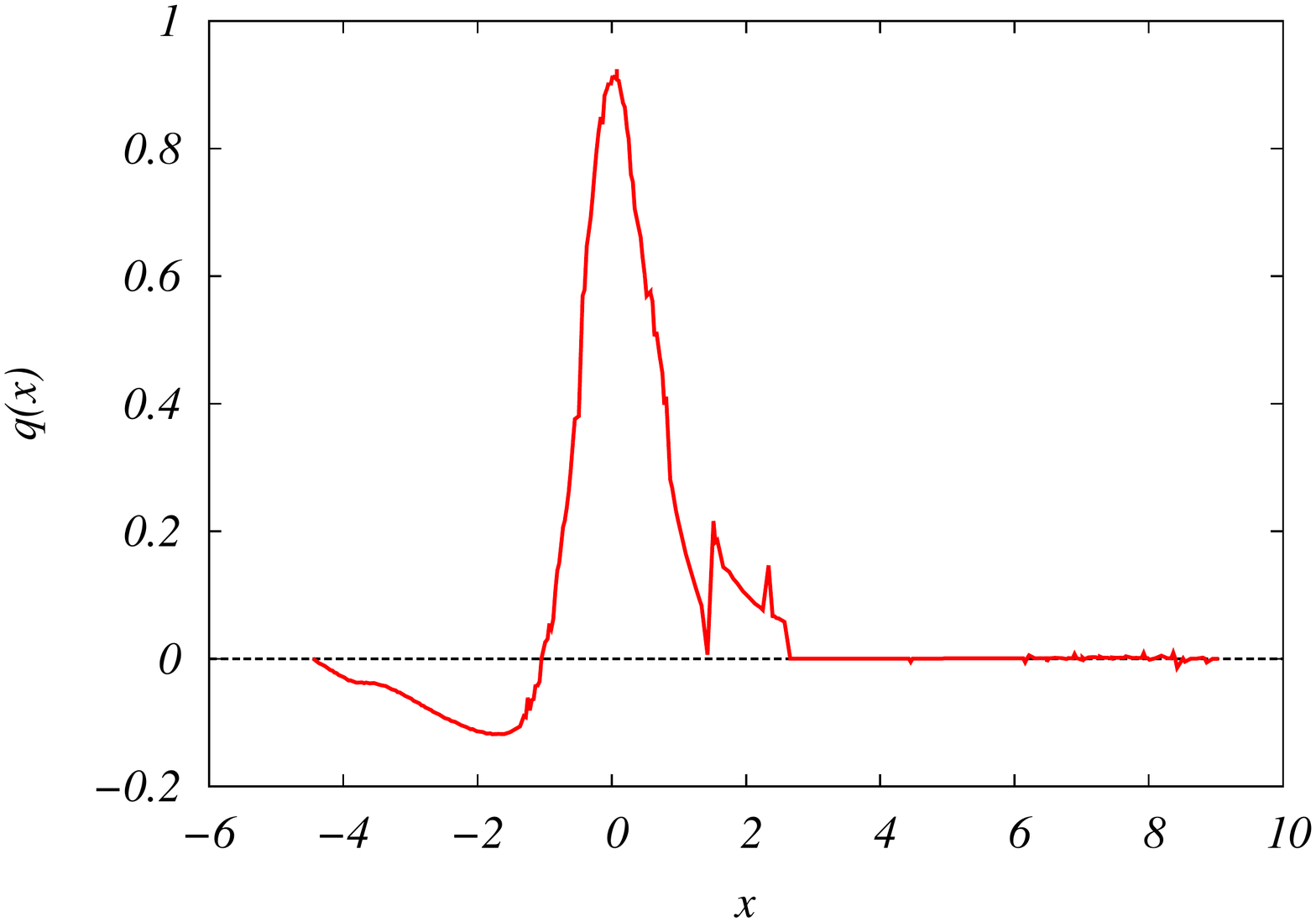}
\end{center}
  \caption{(Color on-line) Absorption line obtained from the ESR lineshape B4
latosol sediment of Balan \etal~\cite{Balan} fig.2. The negative wing of the pdf is taken care of 
by upward shifting of the curve.}
\label{BalanB4}
\end{figure}

The age is determined from the $R_{10}$ evaluation obtained as 
$R_{10}$=~-0.575 with age=1.932  Gyr
Using the extrapolated procedure outlined in the main text, we get an age= 54.75 Mega-years (Myr). Lorentz and Gauss CSD ages are 8.51 Myr and 11.78  Myr  respectively (cf. Table~\ref{Ages}) whereas Balan \etal~\cite{Balan} estimated it around 23.9 Myr.

The third example is drawn from Jonas~\cite{Jonas} who worked 
extensively on fossil tooth enamel dating (spanning several hundred thousand 
years up to 2 Myr).

\begin{figure}[htbp]
\begin{center}
\includegraphics[angle=0,width=100mm,clip=]{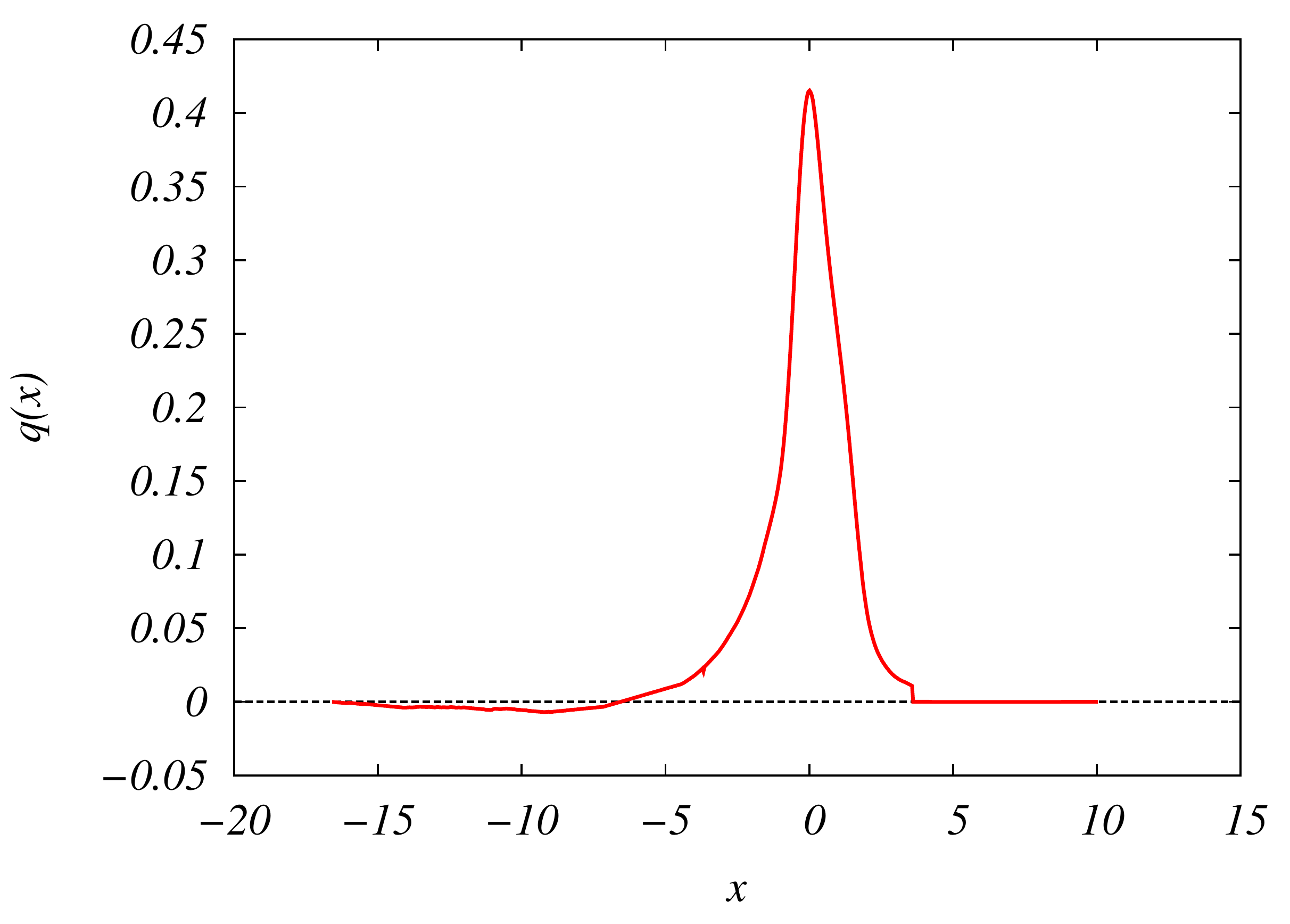}
\end{center}
  \caption{(Color on-line) absorption line obtained from the fossil tooth
enamel ESR lineshape of Jonas~\cite{Jonas} fig.7. The negative part of the pdf is treated as in 
the B4 sediment case.}
\label{Jonas}
\end{figure}

From the ESR lineshape (calibrated spectrum of fig.7), we extract the absorption line according 
to the above protocol with the result displayed in fig.~\ref{Jonas}.

We find the CSD between a Lorentzian (resp. Gaussian) $p(x)$ and 
Jonas~\cite{Jonas} based $q(x)$ as
1.49 $\times 10^{-2}$  and  1.17 $\times 10^{-2}$  
yielding respective ages 7.29  Kyr  and 14.47  Kyr.

For the sake of comparison, if we use the $R_{10}$= -3.26 value (not to be used for 
the tooth enamel era but for ancient carbonaceous matter as in 
Bourbin~\etal~\cite{Bourbin}), 
we get an age of 3.84 Gyr, a value larger than 3.5 Gyr the maximum accepted age of 
organic matter. On the other hand, if we use the interpolation method we get 97.08 Myr 
which is beyond the 2 Myr limiting age of fossil tooth enamel.

\end{document}